\documentclass[3p,times]{elsarticle}
\usepackage{graphicx}
\usepackage{amssymb}
\usepackage{soul}
\journal{\textsc{arXiv}}

\usepackage{amsmath,amsfonts,amssymb,graphicx,color,times,bbm,psfrag,dsfont,slashed,bigints,bm}
\newcommand{\be}{\begin{equation}}
\newcommand{\ee}{\end{equation}}

\definecolor{darkblue}{rgb}{0,0.1,0.5}
\definecolor{darkred}{cmyk}{0,1,1,0.55}

\usepackage{hyperref}
\hypersetup{colorlinks,bookmarksopen,bookmarksnumbered,
linkcolor=darkblue,pdfstartview=FitH,urlcolor=black,citecolor=OliveGreen}

\begin{document}
\begin{frontmatter}

\title{A new approach to dark matter from the mass-radius diagram of the Universe}
%\title{Elsevier \LaTeX\ template\tnoteref{mytitlenote}}
%\tnotetext[mytitlenote]{Fully documented templates are available in the elsarticle package on \href{http://www.ctan.org/tex-archive/macros/latex/contrib/elsarticle}{CTAN}.}

%% Group authors per affiliation:
\author{Aldo Ianni\fnref{myfootnote1}} 
\author{Massimo Mannarelli\fnref{myfootnote2}}
\author{Nicola Rossi\fnref{myfootnote3}}
\address{Laboratori Nazionali del Gran Sasso (INFN), Via G. Acitelli, 22 67100 Assergi L'Aquila, Italy}
\fntext[myfootnote1]{\texttt{aldo.ianni@lngs.infn.it}}
\fntext[myfootnote2]{\texttt{massimo.mannarelli@lngs.infn.it}}
\fntext[myfootnote3]{\texttt{nicola.rossi@lngs.infn.it}}

%% or include affiliations in footnotes:
%\author[mymainaddress,mysecondaryaddress]{Elsevier Inc}
%\ead[url]{www.elsevier.com}

%\author[mysecondaryaddress]{Global Customer Service\corref{mycorrespondingauthor}}
%\cortext[mycorrespondingauthor]{Corresponding author}
%\ead{support@elsevier.com}

%\address[mymainaddress]{1600 John F Kennedy Boulevard, Philadelphia}
%\address[mysecondaryaddress]{360 Park Avenue South, New York}

\begin{abstract}
Modern cosmology successfully deals with the origin and the evolution of the Universe at large scales, 
but it is unable to completely answer the question about the nature of the fundamental objects that it is describing. As a matter of fact, about  $9$5\% of the constituents of the Universe is indeed completely unknown: it cannot be described in terms of known particles. Despite  intense efforts to shed light on this ``literal" darkness by dark matter and dark energy direct and indirect searches, not much progress has been made so far. In this work, we take a different perspective  by reviewing and elaborating an old idea of studying the mass-radius distribution of structures in the Universe in relationship with the fundamental forces acting on them. As we will describe in detail, the distribution of the observed structures in the Universe is  not completely random, but it reflects the intimate features of the involved particles and the nature of the fundamental interactions at play. The observed structures cluster in restricted regions of the mass-radius diagram linked to known particles, with the remarkable exception of very large structures that seem to be linked to an unknown  particle in the sub-eV mass range.
We conjecture that this new particle is  a  self-interacting  dark matter candidate.
\end{abstract}


\end{frontmatter}

\section{Introduction}
What we know about the Universe comes from its \emph{visible} matter component,  
accurately described by the Standard Model of particle  physics and by General Relativity (GR). However, 
large-scale structures
show a compelling evidence for a missing hidden component: only about $5\%$ of the Universe energy content seems to be visible.  The unknown components can be  divided, for their physical properties, in \emph{dark energy}, representing about $70$\% of the energy budget of the Universe, and \emph{dark matter} (DM),  contributing for about the $25$\%  (for reviews see for instance~\cite{bib:sola, bib:bertone}). The first is an unknown form of energy that affects the Universe dynamics at the largest scales. Its existence is  mainly inferred from Type Ia supernovae luminosity distance against red-shift measurements, showing an acceleration in the Universe expansion~\cite{Clocchiatti2006}. The latter is a form of matter accounting for about $85\%$ of the matter in the Universe. In contrast with the visible component, it does not absorb, reflect or emit electromagnetic radiation (or at least those processes are extremely suppressed), and
therefore in this sense it is named ``dark". 
Although  DM is invisible to the observations based on electromagnetic interactions, it is anyway  evident by several gravitational effects. Assuming the validity of  GR, with its weak field Newtonian force approximation, 
 the DM seems to dominate the dynamics of large gravitating objects, from small galaxies to galaxy clusters, and the structure formation since the early stages of the post Big Bang epoch.  Furthermore, the presence of this  hidden form of matter is required to explain various astrophysical observations including the cosmic microwave background (CMB) anisotropies~\cite{Chen:2003gz,Pierpaoli:2003rz,Padmanabhan:2005es}, the gravitational lensing produced by galaxies and cluster of galaxies~\cite{Massey:2010hh} as well as the dynamics of group of galaxies and many other indirect observations, see~\cite{Gaskins:2016cha} for a review. 

About the particle nature of the DM, many candidates have been proposed in years, often associated to extensions of the Standard Model of particle physics, since the model itself does not include any suitable solution. Proposed candidates range from very light DM with mass $\lesssim \mu$eV, as axions~\cite{Peccei:1977hh, Peccei:1977ur,Preskill:1982cy,Chadha-Day:2021szb} and axion-like particles (see~\cite{Choi:2020rgn} for a recent review), to primordial black holes (BHs)~\cite{1966AZh....43..758Z, 10.1093/mnras/152.1.75} with mass up to $10^{21}$ g (see~\cite{Villanueva-Domingo:2021spv} for a review), passing through sterile neutrinos in the keV--MeV mass range~\cite{Lesgourgues:2006nd, Boyarsky:2018tvu}, and  the popular weakly-interacting massive particles (WIMPs) with mass in the range $\sim 1-10^4$\,GeV \cite{bib:bertone}. The WIMP hypothesis has been for a long time the preferred paradigm, because WIMPs  emerge naturally from  extensions of the Standard Model, as in the theoretical framework of Supersymmetry~\cite{Jungman:1995df}.

Unfortunately,  the identity of the DM particle is elusive to both indirect and direct searches. On the direct search side, longstanding attempts of experimental detection through dedicated techniques in underground laboratories or looking for hidden channel processes in relativistic accelerators have only produced  null results (for a review see for instance \cite{ParticleDataGroup:2020ssz}). Being mainly based on gravitational effects, the astrophysical as well as the cosmological observations  can hardly help to  distinguish between the various DM candidates.  In addition, one of the most popular cosmological model, known as the $\Lambda$CDM model,  is recently facing a number of challenges at scales smaller than $\sim$\,1\,Mpc \cite{bib:bullock}. 
Therefore,  even if its gravitational effects are deeply  investigated, the particle identity of DM  is still unknown.

Instead of focusing on a particular DM model, we propose a different approach to the dark matter problem based on an old idea of  Weisskopf  ~\cite{bib:Weisskopf} and Carr and Rees~\cite{bib:Carr} on how one can characterize the distribution of known \emph{structures} in the Universe in relationship with the fundamental forces acting on them. In particular, we analyze the typical mass and length scales of several structures, ranging from nucleons to clusters of galaxies, showing that their distribution in the mass-radius diagram is not randomly scattered, but clustered according to the fundamental interactions at play. Moreover, structures  describing different objects characterized by the same type of interaction seem to be related  by a simple rule. Remarkably, we will show how by the clusterization of macroscopic structures and by building the corresponding connections, one can infer the approximate masses of the proton and of the electron. These mass scales  emerge as those of the basic constituents of ordinary matter. Upon extending the same approach to large-scale structures, we put forward the hypothesis of a new mass scale in the sub-eV range, which could be related to the  DM sector.

Our work is organized as follows.
In Sec.~\ref{sec:obj}  we discuss the existence of relationships between various objects in the Universe  determined by the interactions between their fundamental components. In Sec.~\ref{sec:mass} we present the general properties  of the the mass-radius diagram.  We discuss the extended structures of the Universe in   Sec.~\ref{sec:extended} and their relations to  the microphysics. In particular, we identify the domain in which each fundamental interaction is dominant.
In Sec.~\ref{sec:dark} we dwell on  very large structures and on how a very small mass scale emerges by a simple extrapolation procedure. We also present a simple model that approximately reproduces the mass-radius relation of galaxies. In Sec. \ref{sec:rotation} we use  the matter distribution determined by our simple model to construct approximate  rotation curves for galaxies.    Employing as fitting parameters the halo mass and radius, we obtain rotation curves in very good agreement with observations.  We draw our conclusions in Sec.~\ref{sec:conclusions}.
\section{Structures and their parameters} 
\label{sec:obj}
The classification of all objects in the Universe, from subatomic particles to the Universe itself, can be realized by looking at some  suitable physical observables in a multidimensional parameters space. In general,  we expect that the distribution of the observable objects is not random, because any physical object is realized  by the interaction of elementary constituents. In other words, we expect that objects having the same constituents, and whose equilibrium is determined by the same fundamental forces, form clusters that are related among them by some simple rule.  Within a cluster the various elements can be distinguished by some parameter, however, if one is just interested in the relationship between clusters, such a distinction is unnecessary. 
The famous Hertzsprung-Russell diagram \cite{bib:textbook}, where magnitude versus temperature of stars is reported, is a neat example of this sort. The clustered structures in it help astrophysicists to understand the stellar evolution, inspiring revolutionary intuitions.

At the microscopic level an example of  clusterization is given by atoms. One may classify the chemical elements by using various parameters, for instance their atomic number and some  chemical property, such as their ionisation energy,   as in  the Mendeleev table  of elements. This classification is certainly useful in chemistry, however, in our view the most important aspect is that the periodicity of the table reveals the existence of more elementary objects, namely that atoms are made of interacting nucleons and electrons.
For our purposes  it is indeed sufficient  to classify the chemical elements as a  cluster of objects made of nucleons and electrons interacting through the electromagnetic force. We can then characterize the whole cluster by  atomic mass and radius. The atomic  mass  ranges between $m_p$ and about $300 m_p$,   where $m_p$ is the proton mass, while the radius, of order  $10^{-8}$cm,  is dictated by the strength of the electromagnetic interaction.  
A second relevant example is given by hadrons:   the fundamental as well as the  resonant states can be classified  for instance according to their mass and spin. Also in this case the periodicity of the hadronic states  reveals  the existence of a substructure. In particular, baryons can be described as the  arrangement of three valence quarks whose interactions  mediated by gluons is described by the Quantum Chromodynamics (QCD), see for instance \cite{Halzen-Martin, Cheng-Li}. Using mass and size, we have that all known baryons have a mass between $m_p$ and about $6 m_p$ (for bottom baryons) and size of about $10^{-13}$cm, which is determined by the strength of the strong interaction.  
Therefore, if one looks at the masses and typical length scales of atoms and  baryons, one can easily see that they form two distinct clusters. With logarithmic accuracy, atoms cluster at a mass of order  $10-100$\,m$_{\rm p}$, and radius of order 10$^{-8}$\,cm, while baryons have masses of the order of the proton mass and radii of order 10$^{-13}$\,cm.  The most striking difference is the typical length scale characterizing each cluster, indeed  both atoms and baryons have a mass that is dictated by the proton mass, therefore their masses are close, while the typical sizes are different: the atomic size is determined by the electromagnetic interaction, while the baryonic size is determined by the strong interaction. Note that within each cluster, one can distinguish the different elements by using a different parameter. For example, chemical elements can be distinguished by their ionisation energy,  which does not make any sense for baryons.  Since in our approach we  deal with clusters and not with the single elements of the clusters, we will  assume a coarse grained description of matter in which only the mass and size matter, while any \emph{internal} parameter differing from mass and size has been \emph{integrated out}. 

For macroscopic structures one finds that clustering in terms of mass and size still holds. 
As an example, neutron stars have masses of about $1.4 M_\odot$ and radii of about $ 10$ km, while white dwarfs have similar masses, on a logarithmic scale, but  radii of order $10^3$ km. Therefore,  neutron stars and white dwarfs form two distinct clusters of objects with similar masses but different radii. As in the previous example, the basic difference between the size of the elements of the two clusters  is determined by the different interaction at play. In the first case, it is the nucleon degeneracy pressure and the nuclear forces that determine the hydrostatic equlibrium, while in the latter it is the electron degeneracy pressure that plays the same role. 

This approach can be extended to understand the clustering of all objects in the Universe, by using the simplest parameters in common to all objects, namely their mass and radius (size). Extending the argument discussed above, any other parameter, as spin, or charge that can be used to distinguish the elements within each cluster becomes  irrelevant in the coarse grained description  we perform.  Regarding the useful  parameters for  clusterizing objects, we note that  parameters as temperature, or luminosity, cannot be defined for atomic and subatomic structures and in any case they depend on the kinematic state of the object;   therefore, they are not universally comparable.  
For these reasons we investigate exclusively the mass-radius parameter space. 

We have already given a few examples of clustered structures, however one may think that an exhaustive  classification of all objects in the Universe by means of mass and radius would result in a completely random distribution. As we will see,   this is not the case. Clustered structures, with clear connection pattern,  unequivocally  appear, while wide regions of the mass-radius parameter space remain completely empty. In other words, not all objects with whatever mass and radius are realized in the Universe, due to the
existence of a restricted number of fundamental constituents,  fundamental interactions, and formation mechanisms.
 
 Far from believing that the existence of clustered structures is a  mere coincidence, it is clear that
intrinsic rules are at play determining allowed regions and  forbidden areas. However, the reason why some regions are populated could be also contingent on a physical process that formed an object with a specific radius and mass. In other words, the reason why an object does not exist could be the absence of a production mechanism rather than the instability of the object itself.

This reasoning does not exclude, from a philosophical point of view, a drift towards some radical view as the \emph{anthropic principle}, where this kind of arguments actually arose at first~\cite{bib:Barrow}. 
Nevertheless, the use that we do of the mass-radius diagram is rather pragmatic: we try to use it to connect the macro- and the micro-physics with the purpose of shedding light on the origin of the missing mass of the Universe.

\section{The mass-radius diagram}
\label{sec:mass}
 Weisskopf  in 1975~\cite{bib:Weisskopf} and  Carr and Rees in 1979~\cite{bib:Carr} have shown how one can  semi-quantitatively determine the properties of very different structures starting from the masses of fundamental particles and the  strengths of the fundamental interactions. Their aim  was not to provide a detailed description of nature, but to show how the typical  properties of a number of  objects emerge by a wise use of fundamental parameters. In the present study we build on their works showing how the masses and the length scales  of many structures present in the Universe depend on fundamental constants. 
As in \cite{bib:Weisskopf, bib:Carr}, the goal is not to provide an accurate description of  structures; therefore, an approximation of one order of magnitude is considered acceptable. 

\begin{figure}[htb!]
\centering
\includegraphics[width=1.1\textwidth]{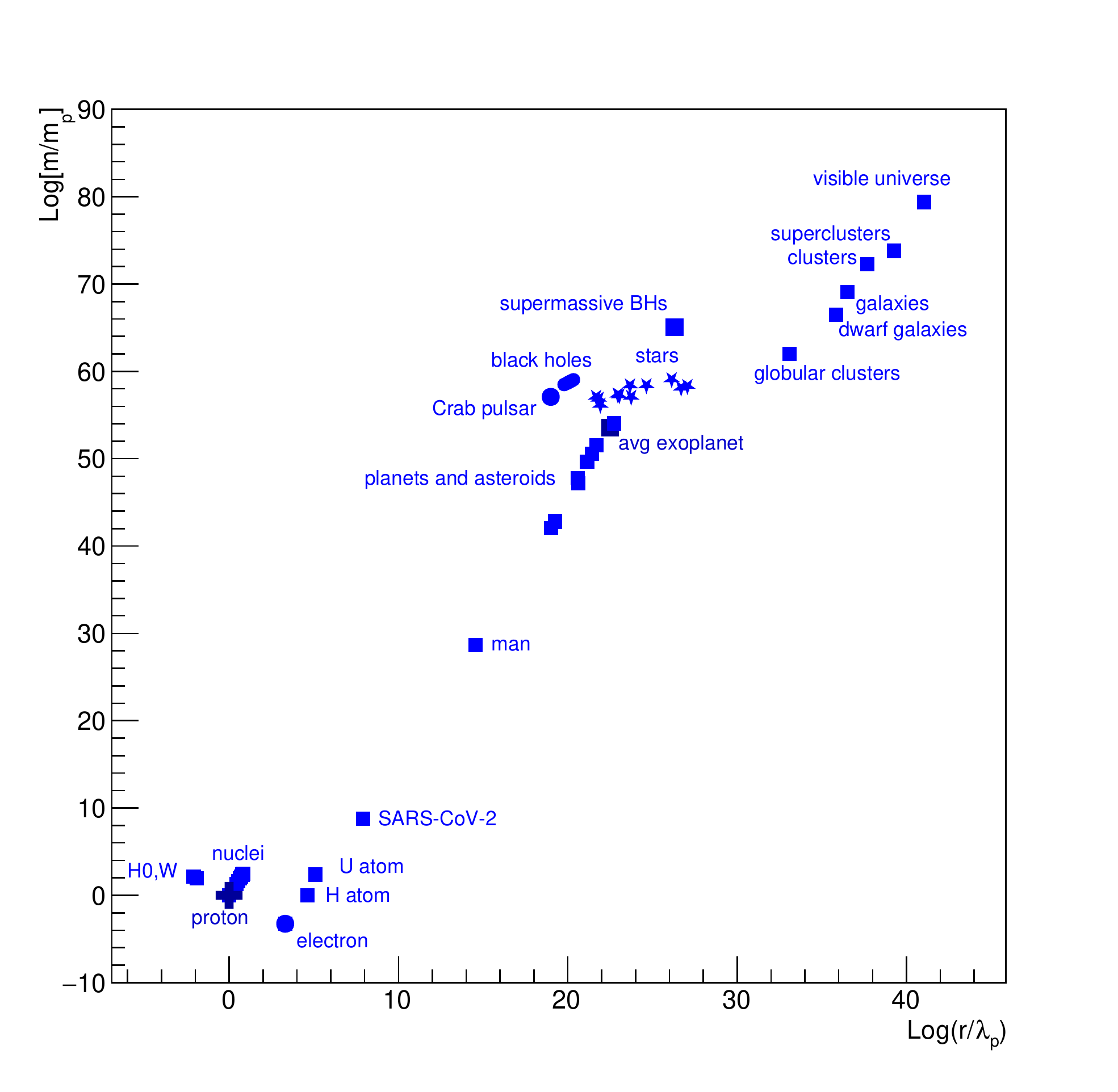}
\caption{Logarithmic plot of mass and radius  of different structures present in the Universe normalized to the proton mass and to the Compton wavelength of the proton.
For the fundamental particles the radius is given by the  corresponding Compton wavelength. For some structures, as nuclei, atoms, planets and stars,  we use the measured radius. In few cases, as for the point marked as ``man", we use   some typical length scale. For very  large structures we  report the average values of virialized mass and radius.} 
\label{fig:1} 
\end{figure}

In Fig.~\ref{fig:1} we present a double logarithmic mass-length diagram showing a number of representative structures of the Universe. The use of a log-log scale is due to the fact that we are not interested in the details of macroscopic structures.  
In the diagram,  we normalize masses to the proton mass, $m_p$, as in~\cite{bib:Carr}, and the radius to the   Compton wavelength of the proton 
\be 
\lambda_p = \frac{h}{m_p c} \simeq 1.3\times 10^{-13}\, \text{cm} \label{eq:lambda_p}\,,
\ee 
with $h$ the Planck's constant and $c$ the  light velocity in vacuum.
The reason  behind  this particular normalization is two-fold. First, we know that most of the matter in the visible Universe consists of baryons, therefore using the proton mass and the typical proton wavelength is the most natural choice. Second, we are not interested in the subatomic physics, meaning that any  length scale below about $1$ fermi  is assumed to be integrated out.  For this reason  in our analysis we will not consider exotic macroscopic objects emerging directly from QCD, such as strange stars~\cite{Alcock:1986hz} or pion stars~\cite{Carignano:2016lxe, Brandt:2017zck, Andersen:2018nzq, Mannarelli:2019hgn}. 

As it is shown in Fig.~\ref{fig:1}, the structures of the Universe occupy only a few regions of the mass-length diagram, showing interesting patterns: they tend to cluster in particular regions  and  the clusters seem to be simply connected by straight lines. 
More in detail, in Fig.~\ref{fig:1} we report a sample of $24$ BHs from the recent  Ligo-Virgo Catalogue~ \cite{bib:Ligo-Virgo} as well as a representative point of supermassive BHs~\cite{King:2003ix, Willott_2003}; the Crab Pulsar (neutron star); a number of asteroids and planets from our solar system: Deimos, Phobos, 4 Vesta, Ceres, the Moon, the Earth, Mars, and Jupiter; an  exoplanet representation averaging a sample of about 2000 observations of available data \cite{Akeson2013} (the size of the point approximately corresponds to one sigma of the distribution in mass and radius); 
a few stars ranging from white dwarfs to red supergiants: Procyon B, Sirius B, Proxima Centauri, Altair, Procyon A, Zeta Ophiuchi, Sun, Rigel, Eta Carinae,  Betelgeuse, and Canis Majoris;  globular  clusters of stars, for them we use a geometric average of the virial mass and tidal radius for $163$ galactic clusters \cite{urlgcstars}; dwarf galaxies, galaxies, cluster of galaxies, and super-clusters of galaxies, for them  we use the \emph{average} (virial) mass and size from  \cite{Revaz2018, NBahcall1999}; few nuclei up to the uranium nucleus; the hydrogen and the uranium atoms.    
Finally, subatomic particles occupy the left bottom edge of the diagram. In particular,  we have  reported the points corresponding to the proton and to the electron. The electron is a point-like particle, therefore we have used as representative length scale its Compton wavelength. Actually any elementary particle can be considered  a point-like object; however, as far as we consider its interactions, the smallest meaningful quantum mechanical (QM) length scale corresponds to its Compton wavelength.  

The  points shown in Fig.~\ref{fig:1}  do not clearly represent an exhaustive list of all the existing objects; however,  as far as we are interested in understanding the fundamental properties of matter,  they are a representative sample of all   the observed objects in the Universe. As we will discuss and better clarify in the following, the objects that we have not reported in Fig.~\ref{fig:1} would cluster close to the  points already depicted in the same figure  or would lie close to the lines joining clusters characterized by the same interaction. This is a consequence of the fact that the structures in the Universe cannot have arbitrary mass and radius.

% - - - - - - - - - - - - - - - - - - - - - - - - - - -
\subsection{Boundaries}
% - - - - - - - - - - - - - - - - - - - - - - - - - - -
We first note that there exist natural boundaries in the mass-radius diagram. A left boundary is determined by GR and it is named in Fig.~\ref{fig:2} the ``black hole" boundary. It is given  by
\be \label{eq:BHboundary}
R_\text{Sch}(m) = \frac{2 G_N m}{c^2}\,,
\ee
where $G_N$ is the gravitational constant. This boundary simply implies that  any structure of mass $m$ must have  a radius equal or larger than the Schwarzschild radius, $R_\text{Sch}(m)$.  Clearly, the selected BHs from the Ligo-Virgo Catalogue \cite{bib:Ligo-Virgo} as well as the shown supermassive BHs~\cite{King:2003ix, Willott_2003} are all grouped on the BH boundary line at large masses. Any other BH that we have not reported in the mass-radius diagram  would be on this line. Notice that  we terminate the BH boundary line at  
the Planck's point  $(2 r_\text{Pl},M_\text{Pl})$ corresponding to a state with a mass equal to the Planck's mass, $ M_\text{Pl}$, and a radius equal to the corresponding Schwarzschild radius. The Planck's mass and radius are the only mass and length scales that can be built from the fundamental constants  $G_N$, $c$ and $\hbar = h/2\pi$ and   are respectively given by
\begin{align}
\label{eq:planck}
M_\text{Pl} & =\sqrt{\frac{\hbar c}{ G_N}}\simeq  2.2 \times 10^{-8}\,\text{ kg}\,, \\
r_\text{Pl} & = \sqrt{\frac{G \hbar}{c^3}}\simeq 1.6\times10^{-33}\,\text{ cm}\,.
\end{align}

The Planck's mass is the largest fundamental mass scale, meaning that any known particle has a smaller mass, therefore  $(2 r_\text{Pl},M_\text{Pl})$ defines a limiting  point on the mass-radius diagram, as shown in Fig.~\ref{fig:2}. 
 
A second  boundary can be defined requiring the localisability of elementary particles, corresponding to a length scale equal to the Compton wavelength 
\be\label{eq:QM_boundary}
R_\text{C}(m) = \frac{h}{m c}\,,
\ee 
and 
it is named the ``quantum mechanics" boundary in Fig. \ref{fig:2}. 
Any particle, elementary as the  electron, or composite as the proton, lies on the QM boundary. For instance, the electron can be identified with the point: $(r_e,m_e)$, 
where 
\be\label{eq:mere}
m_e c^2 \simeq 0.511 \text{ MeV} \qquad \text{and} \qquad  r_e = \frac{h}{m_e c} \simeq 2.4 \times 10^{-11} \text{cm}\,,
\ee are the mass and  the Compton wavelength of the electron, respectively. For extended structures, instead of the Compton wavelength we use the radius of the object. In this way, any object, from elementary   particles to   galaxies can be placed  within the BH and QM boundaries. %
Note that the two boundaries intersect at $R= \sqrt{2} R_\text{Pl}$, corresponding to $M=  M_\text{Pl}/\sqrt{2}$.

% - - - - - - - - - - - - - - - - - - - -
\section{Extended structures}
\label{sec:extended}
% - - - - - - - - - - - - - - - - - - - -
We will now dwell on the extended structures present in the mass-radius diagram.  
Since we are looking for a connection between structures and fundamental interactions, let us first characterize the strength of the gravitational, electromagnetic,  weak and strong  interactions by means of the corresponding fine structure  constants, respectively given by
\begin{align} \alpha_G &= \frac{G_N m_p^2}{\hbar c} \simeq 5.9\times10^{-39}\label{eq:alpha_G}\,,\\
\alpha_e &= \frac{e^2}{\hbar c} \simeq 7.3\times10^{-3} \label{eq:alpha_e}\,, \\
\alpha_W &= \frac{G_F m_e^2}{(c\hbar)^3} \simeq 3.0\times10^{-6}\,, \\  
\alpha_s & \simeq 0.1 \label{eq:alpha_s}\,,
\end{align}
where $e$ is the electron charge and  $G_F$ is the electroweak coupling. Some comments are in order. The  weak interactions are mainly related with  the time evolution of matter, for instance they determine the decay of some heavy elements or  the cooling properties of  stars. For this reason they will not play a relevant role in the following discussion,  we will indeed assume that matter is catalized in  a static or  slowly evolving state. The reported strength of the strong interaction can be obtained by comparing some typical strong and electromagnetic decay lifetimes, see for instance \cite{Halzen-Martin}, and it is therefore meant to be the effective coupling of the interaction mediated by mesons.  More precisely, it corresponds to the strength of the strong interaction  at about the $Z$-boson mass scale, see~\cite{Halzen-Martin,Bethke:2006ac}. 

% - - - - - - - - - - - - - - - - - - - - - - - - - - - - - - - -
\subsection{The gravity domain}
% - - - - - - - - - - - - - - - - - - - - - - - - - - - - - - - - 

Each fundamental interaction plays an important role   in a certain region of the mass-radius diagram. We begin with considering gravity, which is extremely strong  near  the BH boundary; as we have seen this boundary   is indeed completely determined by GR.  
The largest possible structure in which gravity is the only relevant interaction is the whole Universe which corresponds to a limiting point on the mass-radius diagram. The mass  of the visible Universe can be written as $M_U \sim 0.22 m_p \alpha_G^{-1} r_p^{-1} r_U $, while its  radius is $r_U \sim c/H_0$, being $H_0$ the present day Hubble expansion rate, and $M_U \sim c^3 t/6\pi G_N$ with $t\sim 1/H_0$ for a matter-dominated Universe. Hereafter, we will not include the Universe point in our discussions, since it only refers to the visible part of the whole Universe: 
its real size and mass are not observable. 
For structures on the BH boundary, that is for BHs, there is no equilibrium between gravity and any other force:  their compactness is so large that gravity dominates. However, as we move to the right of the BH boundary the compactness decreases and different interactions are at play to contrast  gravity. The most compact objects are  neutron stars, in which the fermionic degeneracy pressure of neutrons and the nuclear short-range repulsive interaction balance the gravitational attraction. However, there exists an empty region between the BH cluster and the neutron star cluster. This is an extremely important point:  in general,  one cannot move with continuity between regions where different interactions are at play. Regarding the BH line, the region very close to this boundary  cannot be populated, indeed objects with a compactness larger than the Buchdal's limit $m/r = 4/9$ collapse to BHs~\cite{Buchdahl:1959zz}. For this reason, the region $4/9 < m/r < 1/2 $ must be empty: it corresponds to one of the  forbidden regions in the mass-radius diagram. These regions can in general be characterized by the fact that it is not possible to realize any sort of equilibrium between the fundamental forces to form  a stable object; in the considered case, there is no equation of state (EoS) of matter that can sustain equilibrium for a structure with $4/9 < m/r < 1/2 $.
A different way of looking at this result is that   the BH line has no smearing: no structure can exist sufficiently close to the BH boundary.
The only exceptions are Kerr BHs, which are slightly deformed by  rotation, see for instance~\cite{Misner:1974qy}.

A second important general feature of the mass-radius diagram is that along lines determined by some specific interaction there may be empty regions.  Along the BH boundary line we have only observed BHs in  restricted regions, roughly around $10-100 M_\odot$, for LIGO-Virgo detected BHs, see for instance~\cite{LIGOScientific:2020ibl}, and $10^8-10^9 M_\odot$ for supermassive BHS, see for instance~\cite{King:2003ix, Willott_2003}. Very light BHs with mass $M \lesssim 10^{15}$ g possibly created during inflation have so far evaporated and it is unclear whether primordial BHs with masses below $M_\odot$ can be created. In any case,  there is a large BH mass range that does not seem to be populated. Presumably, there is no fundamental reason for that: it is possible that there is no way in Nature  of producing  BHs in some mass ranges, or maybe we  will observe these objects in the future. 

As we move to the right of the BH line we find less compact objects, where gravity becomes less important and eventually  becomes irrelevant while  different interactions become dominant. As happening for the  gravitationally dominated structures, we expect the dominant interaction to provide the relevant mass and length scale. 

% - - - - - - - - - - - - - - - - - - - - - - - - - - - - - - - -
\subsection{The electromagnetic  domain}
% - - - - - - - - - - - - - - - - - - - - - - - - - - - - - - - - 

Several structures can  be described by means of the electromagnetic interaction, which depends on the coupling constant $\alpha_e$, see Eq.~\eqref{eq:alpha_e}. The lightest system is the hydrogen atom, with a mass $m_H \simeq m_p$  and  radius   \be r_H \simeq a_0 = \frac{r_e}{2\pi\alpha_e} \simeq 5.3 \times 10^{-8} \text{cm}\,, \ee 
where $a_0$ is the Bohr's radius. 
Heavier atoms have mass of order $\sim A m_p$, where $A$ is the atomic number, while their radius has a non-monotonic dependence on $A$. Atoms have indeed a very peculiar mass-radius relation:   within a group of the periodic table  the radius typically increases with increasing $A$, while it decreases within a period. However, with good  accuracy,  we can say that atoms form a  cluster in the mass-radius diagram with masses 
\be m_\text{atoms} \simeq  1 \div 100\, m_p\,,  \ee 
and radii 
\be r_\text{atoms} \simeq  1 \div 4 \,a_0 \,,\ee 
where the ranges of values are empirically determined.

  Moving to astronomical solid objects, such as planets with spherical symmetry, 
  we take the simplifying assumption that they are made by one single  chemical component with   atomic/molecular weight  $A$.  Since the electromagnetic interaction is dominant, the matter density of the planet 
\be \rho_\text{planet} \sim \frac{m_p A}{(\eta a_0)^3}\,,
\label{eq:rhoplanet}
\ee
is roughly uniform. Here $\eta a_0$ is the effective size for atoms/molecules with $\eta$  a number weakly dependent on $A$,  for instance in \cite{bib:Weisskopf} $\eta \simeq 1.5A^{1/5}$. It follows that 
\be\label{eq:planets_MR}
M \simeq \frac{m_p A}{(\eta a_0)^3} R^3\,,
\ee
therefore we have that 
\be \label{eq:logM_el} 
\log\left(\frac{M}{m_p}\right) \simeq C + 3 \log\left(\frac{R}{\lambda_p} \right) \,, 
\ee with $C$ a constant, corresponding to the dashed green line named ``atomic density" in Fig.~\ref{fig:2} with slope equal to $3$.  We use the constant $C$  as  a fitting  parameter; as we will see below, this parameter is  linked to the microphysics. 

The ``atomic density" line describes very well a number of objects, however it cannot extend to arbitrary large masses. With increasing mass, gravity comes into play and determines the stability of the planet.  Semi-quantitatively,  the electromagnetic  and the gravitational interactions can be characterized by their respective energy scales,
\be \label{eq:scales}
 E_\text{G} \sim \frac{G_N M^2}{R} \qquad \text{and} \qquad  E_\text{e.m.}\sim N \alpha_e^2 m_e c^2\,,
\ee
where $N$ is the number of molecules and $\alpha_e^2 m_e c^2$ is an estimate of the electromagnetic interaction energy per molecule. Since 
\be \label{eq:NAmp}
M \simeq N A m_p\,, 
\ee we obtain  from Eq.~\eqref{eq:planets_MR} that $R \propto N^{1/3}$, meaning that $E_\text{G} \propto A^2 N^{5/3}$, while $E_\text{e.m.} \propto N$. 
 This different scaling with $N$ reflects the fact that gravity is long-ranged, while the electromagnetic interaction is short-ranged.
It follows that with increasing $N$, gravity will eventually win. The equilibrium will be maintained for large $N$ by decreasing $A$, and this is indeed the reason why very massive planets, as Jupiter,  are made of    hydrogen and helium. 

More in detail, assuming that $E_\text{G} \sim E_\text{e.m.} $, upon using \eqref{eq:rhoplanet} and thus
  estimating the number of molecules  as the total volume divided by the effective volume   $N \sim \left(R/\eta a_0\right)^3$,
  we have that the radius and mass of a planet can be written in terms of fundamental constants as
\begin{eqnarray} 
\label{eq:4}
R \sim 1.3 \left(\frac{\alpha_e}{\alpha_G}\right)^{1/2}\frac{a_0}{A}\eta^{3/2}\,,
\\
\label{eq:5}
M \sim 2.2 \left(\frac{\alpha_e}{\alpha_G}\right)^{3/2}\frac{m_p}{A^2}\eta^{3/2}\,,
\end{eqnarray}
where $\alpha_G$ is defined in Eq.~\eqref{eq:alpha_G}. It follows  that in the equilibrium configurations the mass and the radius increase with decreasing $A$. 
For very light elements mass and radius reach a maximum value of order $10^{30}$ g and $10^{10}$ cm respectively,  close to  the Jupiter's mass and radius. In other words, using only the electromagnetic interaction one can  obtain objects in equilibrium with gravity with a maximum mass comparable to that of Jupiter: objects with larger masses must be sustained by the strong interaction or by the quantum degeneracy pressure. Qualitatively, this can also be seen as follows: if one could replace in the previous equations $\alpha_e$ with $\alpha_s$, then one could obtain equilibrium configurations with larger masses and radii.

Turning back to Eqs.~\eqref{eq:4} and \eqref{eq:5}, we can use them to estimate the masses and radii of sufficiently massive planets.  For planets consisting of SiO$_2$ we find that $R \simeq 8\times10^{8}$ cm and
$M\simeq 8.6\times10^{27}$ g. This \emph{standard} rocky planet is shown in Fig.~\ref{fig:2} and it lies indeed on the ``atomic density" dashed green line. The mass and radius of the standard rocky planet are very close to the Earth's mass and radius, namely $M_\oplus = 6\times10^{27}$ g and $R_\oplus = 6.4\times10^{8}$ cm, indeed 
from \cite{bib:Ringwood, bib:Allegre} we can estimate an average $A\sim 33$ for the Earth. In this case from Eqs.~\eqref{eq:4} and \eqref{eq:5} we determine $M \sim 2.4\times10^{27}$ g and $R=1.2\times10^9$ cm. Therefore, such simplistic arguments give a result which is in agreement with measurements within a factor of $10$.

The simple Eq.~\eqref{eq:logM_el}  seems to correspond  to the universal relation between the mass and radius  of any planet: even exoplanets lie along the atomic density line.    However, unlike the BH line, the atomic density line shows some smearing, meaning that  there are objects with similar masses but different radii. We have already mentioned  the peculiar dependence of atoms' radii on the atomic number.  In order to estimate the smearing of the planet distribution,  we follow a reasoning similar to the one presented in \cite{bib:Weisskopf, bib:Carr, bib:Barrow} to estimate how deformed can be a rocky object. In particular, we will determine the maximum height of a mountain by comparing the typical  electromagnetic force with the gravitational force  at the planet's surface.
The surface  gravity of  an object of mass $M$ and radius $R$ is $g=G_N M/R^2$, upon using Eq.~\eqref{eq:planets_MR} and \eqref{eq:NAmp}, we have that
\begin{equation} \label{eq:6}
g_{planet}\simeq \frac{G_N N A m_p}{N^{2/3}(\eta a_0)^3},
\end{equation}
is the surface acceleration due to gravity for a planet composed of $N$ molecules with atomic number $A$. The maximum size of a mountain, $h_\text{max}$, can be estimated by  \cite{bib:Weisskopf}
\begin{equation} \label{eq:7}
    m_p A g_\text{planet} h_\text{max} = \epsilon  \alpha_e^2 m_e c^2,
\end{equation}
where the left hand side corresponds to the gravitational energy and the right hand side to the energy required to avoid that the solid material sinks into the underneath surface. The small parameter $\epsilon \sim 10^{-2}$ gives the scale factor needed to obtain the right order for the energy to break hydrogen or Van der Waals bonds. Solving for $h_\text{max}$, Eq.~\eqref{eq:7} gives
\begin{equation}
    h_\text{max} \sim \epsilon \eta^2 \frac{\alpha_e}{\alpha_G}\frac{a_0}{A^{5/3}}\left(\frac{m_p}{M}\right)^{1/3} \sim 30\div60 \, \text{km}\,,
\end{equation}
where we have used $A=30\div60$ and $M=M_\oplus$ in the numerical estimate. This result shows that we have obtained the right answer for the maximum height of mountains on the Earth within a factor less than $10$. In general, for any irregular solid object,  the above expression of $h_\text{max}$ gives its   maximum size deformation. Asteroids such as Ceres and 4 Vesta are compatible with this limit, as shown in Fig.~\ref{fig:2}.

A remarkable fact is that  the
``atomic density" line  intersects the quantum mechanical boundary at $m\simeq 0.4$ MeV, close to the electron mass. The  mass scale obtained in this way depends on the parameter $C$ of Eq.~\eqref{eq:logM_el}, which is determined by a best fit of the distribution of stars, planets and atoms. In other words, from the astrophysical  knowledge of the masses and radii of stars, planets and  atoms, we can estimate the electron mass.  The reason is  that the size of these structures is dictated by the electromagnetic interaction of electrons, which plays a dominant role. Any other scale, for instance related to the internal structure of nuclei, is irrelevant for the equilibrium properties of these objects.

\begin{figure*}[htb!]
  \centering
  \includegraphics[angle=0,width=1.1\textwidth]{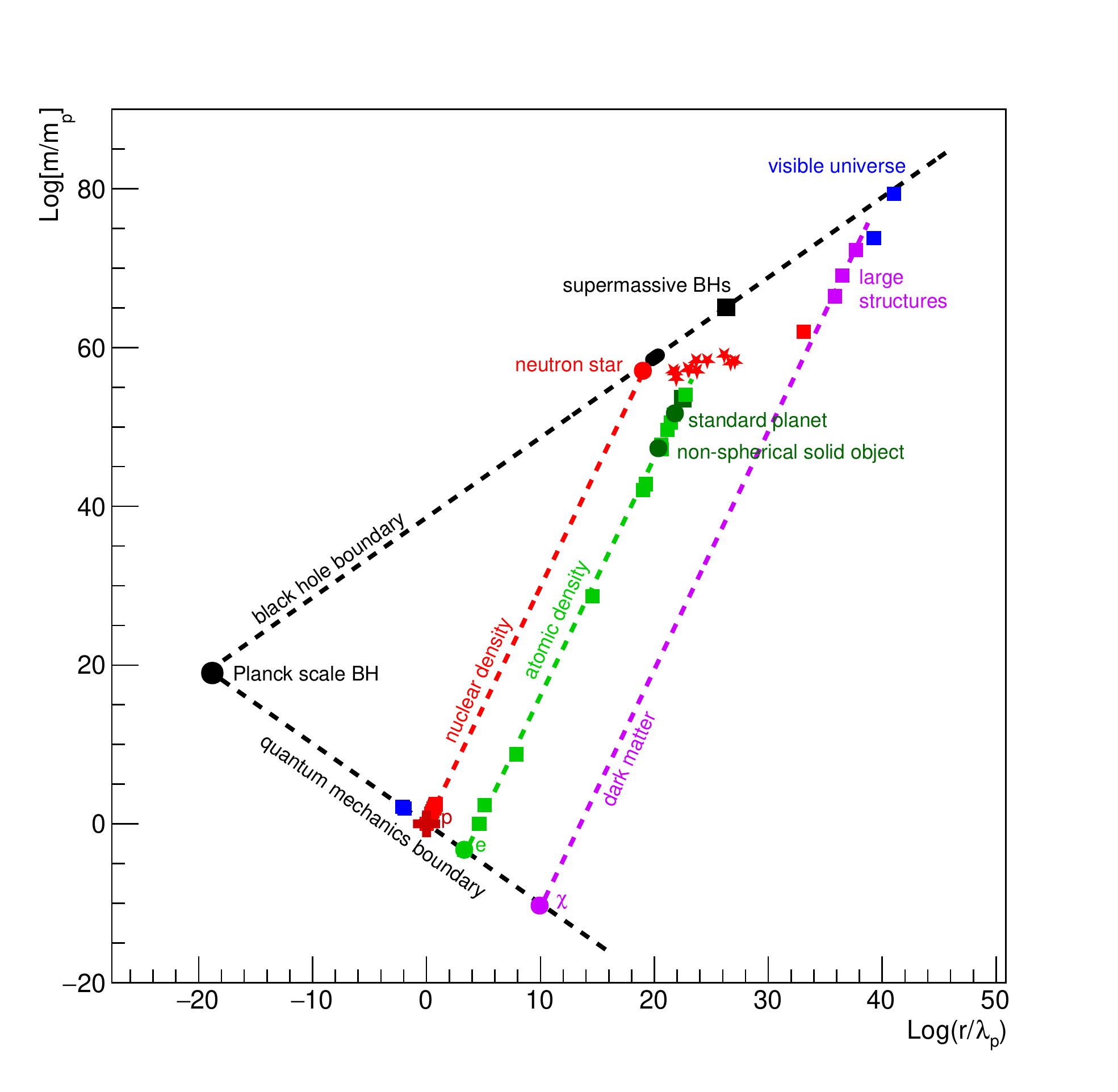}
  \caption{ Logarithmic plot of the  mass and radius   of different structures in the Universe as in Fig.~\ref{fig:1} with boundaries and lines  determined in the present work. The ``black hole" boundary and  the ``quantum mechanics" boundary are  determined by Eqs.~\eqref{eq:BHboundary} and ~\eqref{eq:QM_boundary}, respectively. The ``nuclear density" line connects structures in which the nuclear interaction is dominant. The ``atomic density" line connects structures in which the electromagnetic interaction  dominates.   See the text for more details.} \label{fig:2} 
\end{figure*}

% - - - - - - - - - - - - - - - - - - - - - - - - - - - - - - - -
\subsection{The nuclear interaction domain}
% - - - - - - - - - - - - - - - - - - - - - - - - - - - - - - - -

As we have seen, above Jupiter's mass the gravitational attraction  will eventually become stronger than the  electromagnetic repulsion: nuclear processes must come into play to provide the pressure necessary to maintain equilibrium. Moreover, fermionic matter is so closely packed that the degenerate fermionic pressure becomes sizeable.   The strong interaction between a pair of nucleons is described by the Yukawa potential: $U_s = -\frac{\alpha_s}{r}e^{-r/\lambda_s}$ with $\alpha_s$ the strong interaction coupling constant, see  Eq.~\eqref{eq:alpha_s}, and $\lambda_s \sim 1.4\times 10^{-13}$ cm the typical range of the strong interaction determined by the inverse pion mass.  This length scale is related to the fundamental quantities defined so far by, $\lambda_s \simeq \alpha_s^{-1} \lambda_p/2\pi $, where $\lambda_p$ is the Compton wavelength of the proton, see Eq.\eqref{eq:lambda_p}.

The typical nuclear density can  be written as 
\be
\rho_s \sim \frac{m_p}{\lambda_s^3} \sim 2 \times 10^{14} \text{g cm}^{-3}\,,
\ee
which is of the order of the nuclear saturation density. Since  nuclear matter has a low compressibility, the mass-length relationship involving the nuclear interaction can be written as \be
\label{eq:m_strong} M \sim \left(\frac{\alpha_s}{\lambda_p}\right)^3 m_p R^3\,,
\ee and should roughly characterize any structure in which the nuclear interaction plays a fundamental role. Similarly to the case of structures dominated by the electromagnetic interaction we have that at least for small masses 
\be \label{eq:logM_nucle} 
\log\left(\frac{M}{m_p}\right) \simeq C' + 3 \log\left(\frac{R}{\lambda_p} \right) \,, 
\ee
where $C'$ is a fitting constant. In fact, as can be seen from Fig.~\ref{fig:2},  the $M\propto R^3$ behavior actually  connects nuclei with   neutron stars in the mass-length diagram. Roughly speaking, a  neutron star is  a collection of non-relativistic particles (mostly neutrons) in equilibrium between the inward pressure due to the gravitational attraction and the  outward pressure  due to the    degenerate fermionic pressure (the so-called exchange force) and   the strong repulsive interaction. The short range strong repulsion 
 combined with the quantum exchange forces produce  neutron stars with   masses  $ 1\div2$ M$_\odot$, see for instance~\cite{Misner:1974qy, Shapiro-Teukolsky}. As we will see, if one does not include the short range strong repulsion only neutron stars with masses of the order of $M_\odot$ are stable: more massive neutron stars would collapse.  Although neglecting the strong interaction  is unsatisfactory:  neutron stars with masses $M \simeq 2 M_\odot$ have indeed been observed~\cite{Demorest:2010bx, Antoniadis:2013pzd, NANOGrav:2019jur},  the exchange interaction and the strong interaction are comparable. Therefore, for our semi-quantitative purposes it is sufficient to consider the effect of the exchange forces. In passing, we note that  exchange forces between electrons are the dominant repulsive interaction in white dwarfs, which indeed have masses of about $M_\odot$.
Fermions in a region of volume  $V$  exert a degeneracy pressure 
\begin{equation} \label{eq:9}
    P = \frac{2}{5} \epsilon_F n \left(1 + {\cal O} \left(\frac{T}{\epsilon_F} \right)\right),
\end{equation}
where $n = N/V$ is the average particle density and 
\be
\epsilon_F = \frac{\hbar^2}{2 m} (6 \pi^2 n)^{2/3}\,,
\ee
is the Fermi energy, see for instance~\cite{huang2000statistical}, with $m$ the mass of the fermion.
Since neutron stars are composed by baryons, the Fermi energy is of the order of GeV,  while the temperature few seconds after the neutron star birth is expected to be of the order of keV. For this reasons  we can safely neglect any temperature dependence.  
In this case, the degeneracy pressure can be approximated with  \be P_0=\frac{2}5 \frac{\hbar^2}{2 m} (6 \pi^2)^{2/3} n^{5/3} \,,\ee %being the  size of the region equal to $(V/N)^{1/3}$.
where the $0$ index stands for zero temperature. 
 Neglecting the strong interaction, the spherical  star with mass $M$ and radius $R$ is in hydrostatic equilibrium when 
the fermionic degenerate pressure equals the gravitational pressure, corresponding to \cite{huang2000statistical, bib:textbook}:
\begin{equation} \label{eq:10}
    P_0 = \frac{ \kappa}{4 \pi}\frac{GM^2}{R^4}\,,
\end{equation}
where $\kappa$ is a number of order unity  depending on the mass distribution inside the star. For our numerical estimate we will take $\kappa=3/5$ corresponding to a homogeneous star.
Neglecting any nuclear  and gravitational binding, we can approximate $M \simeq N m_p$ and thus  the relationship in \eqref{eq:10} can be recast to obtain the radius with respect to the mass:
\begin{equation} \label{eq:11}
    R \simeq 3.7 \, \lambda_p \alpha_G^{-1} \left(\frac{M}{m_p}\right)^{-1/3},
\end{equation}
where the numerical prefactor is obtained assuming an homogeneous mass distribution, while $\lambda_p$ and $\alpha_G$ are given in Eqs.~\eqref{eq:lambda_p} and \eqref{eq:alpha_G}, respectively. The above relation shows that for nuclear matter
$M\propto R^3$, which is basically a realization of the fact that   nuclear matter at any density has a very low compressibility.  
Upon substituting  $R\sim10$\,km in Eq.~\eqref{eq:m_strong}
 we obtain $M \simeq M_\odot$, simply meaning that a neutron star can be thought of as a system of closely packed nucleons. 
  The location in the mass-radius plane determined by Eq.~\eqref{eq:11} is reported in Fig.~\ref{fig:2} for $M=M_\odot$, showing indeed that neutron stars and nuclear matter are connected by the dashed red line labeled ``nuclear density" with slope $3$. Extrapolating this line to the quantum mechanical boundary we obtain a mass value $m\simeq 600$ MeV, close to the proton mass. In other words, starting from the observed neutron star and nuclear matter masses and radii we may infer an approximate value  of the proton mass. One important remark is that  nuclear matter and neutron stars are disconnected:  there is no continuous equilibrium  path that connects standard nuclear matter with neutron stars.  neutron stars cannot be built adiabatically piling up neutrons and protons. The reason is that  in  this process one would need to construct nuclei with very large atomic number, which in standard conditions are beta unstable.  To build a neutron star it is indeed needed  a quick and violent compression process, as produced  during a core collapse before the supernova explosion. Nevertheless,  nuclei and neutron stars are in equilibrium in conditions dictated by the nuclear and exchange forces and this is the reason why they are connected by the ``nuclear density" line in Fig.~\ref{fig:2}. 
% - - - - - - - - - - - - - - - - - - - - - - - - - - - - - - - - - - - - - - - -
\subsection{Stars}
\label{sec:stars}
% - - - - - - - - - - - - - - - - - - - - - - - - - - - - - - - - - - - - - - - -
To complete the description of the mass-radius diagram, we now discuss the stellar structures, marked with red star dots in Fig.~\ref{fig:2}. Although approximately lying on the ``atomic density" line, stars have a large smeared distribution, as compared to the other objects: there  exist stars with very different masses and radii.
A semi-quantitative estimate of  such a smearing has been given in~\cite{bib:Carr}; here we briefly report their results.   Stars can be approximated as a  non-relativistic neutral ensemble of $N$ electrons and $N$ protons in which  gravity is balanced by the radiation pressure, the kinetic pressure and the electron degeneracy pressure. Therefore unlike the objects discussed so far in stars different processes are at work: The kinetic energy is provided by nuclear fusion processes, while the electromagnetic interaction determines the scattering of photons on charged particles. The stellar  equilibrium  is therefore determined by a combination of nuclear, electromagnetic, and weak interactions. The latter drives the conversion of protons into neutrons. 
To roughly address this problem one can use the virial theorem, which implies that for the collapsing matter the time-averaged energy of the system equals one-half the time-averaged gravitational energy~\cite{bib:Carr}:
\begin{equation} \label{eq:12}
    T+  \frac{h^2}{2m_e r^2} \sim \frac{1}{2}\frac{G_NMm_p}{R}=\frac{1}{2 r}\alpha_G N^{2/3}\,,
\end{equation}
where  $T$  is the temperature of the collapsing matter  (approximated uniform) and  $R$ is its  radius.
In Eq.~\eqref{eq:12} we have used $R=N^{1/3}r$, being $r$ the average distance between particles, and  we have neglected the degeneracy energy of protons because  $m_p \gg m_e$.   As a function of the particle distance, the temperature reaches the maximum when
$r=2h^2/m_e \alpha_G N^{2/3}$ corresponding to  
\begin{equation} \label{eq:13}
    T_{max} \sim 0.2\, \alpha_G^2 N^{4/3} m_e\,,
\end{equation}
 which should be of the order of  $1$\,keV  to ignite proton-proton fusion.  It follows from Eq. \eqref{eq:13} that the ensemble  of particles becomes a star when $\alpha_G^2 N^{4/3} \geq 0.06\,m_e$, meaning  that $N \geq 0.12\,\alpha_G^{-3/2} \sim 10^{56}$. 
A maximum mass can be obtained by requiring that the star's pressure is not dominated by the radiation pressure;  it is indeed known that any relativistic gas makes stars unstable~\cite{Shapiro-Teukolsky}.   Therefore, at equilibrium the photon gas pressure  $p_\gamma = \frac{\pi^2}{45}T^4$,  is less than the matter pressure, $p_\text{mat} \sim NT$. By equating these quantities one obtains  the upper bound on the maximum number of particles, which reaches $N \sim 32\alpha_G^{-3/2}  \sim 10^{59}$. It follows that the  approximate   stellar mass range is $0.2 \div 100$\,M$_\odot$, while the radii, taking $R=N^{1/3} r$, are within $10^4-10^6$ km. These values are in agreement with 
the distribution of the sample of stars shown in Figs.~\ref{fig:1} and ~\ref{fig:2}, ranging between red dwarfs and  blue giants.

% - - - - - - - - - - - - - - - - - - - - - - - - - - - - - - - - - - - - - - - -
\section{The large-scale structures  and dark matter}
\label{sec:dark}
% - - - - - - - - - - - - - - - - - - - - - - - - - - - - - - - - - - - - - - - -

So far we have not discussed very large structures, such as globular clusters, galaxies,  galaxy clusters and super-clusters. Each of the corresponding points  shown in Fig. \ref{fig:1}  is obtained by  an average over a large sample of such structures.   These points seem to pertain neither to the ``nuclear density"  nor to the ``atomic density" lines, nevertheless they  seem to be aligned. As discussed above, the  ``nuclear density" and ``atomic density" lines are determined by the relevant interactions that operate on protons and electrons, respectively. As we have already noted, one can estimate the mass of the fundamental particles by  the intersection of the  ``nuclear density" and ``atomic density" lines  with the  QM boundary. In this way we have indeed obtained values close to the  proton mass and electron mass, respectively. As shown in Fig.~\ref{fig:2}, a  line named ``dark-matter" can be drawn through the very large structures. It crosses the QM boundary at $r_\chi \simeq 1.6\times 10^{-3}$\,cm and $m_\chi \simeq 10$\,meV. This point, highlighted with the letter $\chi$, could be interpreted as a new fundamental particle relevant for the very large structures. Due to the extrapolation procedure in the log-log scale, the mass cannot be accurately pinned down. We estimate the mass   of this particle to be in the  $1\div100$\,meV range.  More precisely, the intersection of the ``dark-matter" line with the QM boundary in Fig.~\ref{fig:2} represents a lower limit for the mass of a hypothetical DM structure. If DM is an elementary particle, then its mass should be determined by the intersection of the  ``dark-matter" line with the QM boundary, giving a mass in the $1\div100$\,meV range. On the other hand,  if DM is an extended solitonic-like object, its  mass could be larger: the corresponding point   could be anywhere along the ``dark-matter" line of  Fig.~\ref{fig:2}. In the present work we do not discuss the latter possibility; we only elaborate on the existence of a DM elementary particle with mass  $m_\chi \simeq 1\div100$\,meV.

The hypothesis that DM consists of light  particles is not new, for a review see~\cite{Ferreira:2020fam}; it has been put forward with the idea to address open issues on astrophysics and DM distribution. Moreover, it has been proposed as a means to improve the agreement between observations and theoretical calculations of white dwarfs luminosity function, see for instance~\cite{bib:isern}. However, we point out that our argument disfavors ultra-light particles, say below $1$ meV. Although the nature of this DM particle is immaterial for our semi-quantitative discussion, we note that such low-mass particle  can hardly  be a fermion~\cite{Tremaine:1979we}.

In our approach, the low mass scale of the DM particle does not arise from the need of solving any astrophysical problem. It  arises from 
an  extrapolation of the observed very large structures to the quantum level in analogy with the atomic and nuclear cases.   The slope of the ``dark-matter" line in the log-log plot is approximately $3$ meaning that this hypotetical state of matter should have  a low compressibility. This is exactly the same behavior that we found for atomic and nuclear matter and that was based on the fact that there are regions in which one  fundamental interaction is dominant and sufficiently strong to force the $M \propto R^3$ behavior.      To be  more precise, the best fit of the slope of the ``dark matter" line obtained fitting the large structures with the exception of superclusters and globular clusters is about $3.1$. Upon including these two structures the best fit has a smaller slope, of about $2$. 
However, we believe that a different dynamics characterizes these structures. The dynamics of superclusters is related to the Hubble expansion and that of stars clusters to the initial process of galaxy formation. Therefore, we may interpret  superclusters and globular clusters as structures in which various mechanisms are responsible of their mass and radius. Similarly to what we  discussed in Sec.~\ref{sec:stars} for  stars, in this case a simple power law relation between mass and radius  is not sufficient to characterize the whole ensemble of large structures. For this reason we focus on structures which, we believe, are in a steady-state dynamic phase dictated by a single interaction. In other words, the extrapolation from very large structures to the quantum boundary reported here is to be interpreted as follows: it is possible  to infer the value of the DM mass only for  the large astrophysical structures that are dominated by DM, being the latter at least a factor $5$ in average bigger than the baryonic matter. In this case, indeed,  their mass-radius relationship is basically dominated by the DM mass, self-interactions and gravitational interaction, similarly to structures determined by the electromagnetic and nuclear interactions.  

Alternatively, one may adopt a slightly different perspective. Irrespective of any consideration on the fundamental forces at work, from the analysis of the distribution of points in Fig.~\ref{fig:1},  one can infer that nuclear matter and neutron stars are connected by the $\log(m) = C + 3 \log(r) $  law, while   atomic matter, planets and stars are connected by the $\log(m) = C' + 3 \log(r) $  law,  where $C$ and $C'$ are two different parameters which are  linked to the masses of the fundamental microscopic particle  that determines these structures.  These lines are reported in  Fig.~\ref{fig:2} with the labels ``nuclear density" and ``atomic density";  they intersect the quantum mechanics boundary at the mass scales $\sim 600$ MeV and  $\sim 0.4$ MeV, close to the proton mass and to the electron mass, respectively. Some smearing around these lines  can be related to the fact that besides their masses and radii the various  structures reported in Figs.~\ref{fig:1} and \ref{fig:2}  are determined by  some internal parameters. If one assumes a similar reasoning for  very large structures, meaning that they
are described by the $\log(m) = C'' + 3 \log(r) $  law,  one finds that one can choose the parameter $C''$ to appropriately describe the average values of dwarf galaxies, galaxies  and clusters of galaxies. Globular clusters and superclusters are objects  slightly off  this line, therefore some internal parameter determines their deviation from the ``dark matter" line. The parameter $C''$ can then be linked to the mass of the   fundamental microscopic particle: the intersection of the ``dark matter" line with the quantum mechanics boundary gives a mass of the order of  $m_\chi = 10$\,meV.

DM candidates with mass $\lesssim 1$ eV belongs to the so-called \emph{ultra-light DM} models. If not interacting, such a kind of matter would become non relativistic too late after the Big Bang, making it impossible the structure formation according to the well established $\Lambda$CMD paradigm. In some way, those particles should be produced already ``cold'' through self-interaction mechanisms, as e.g. for the misalignment of axions. The self-interaction is a property already implicit in our argument, as we will see in details in the following subsections.

% - - - - - - - - - - - - -
\subsection{The   dark matter density profile in galaxies}
\label{sec:profile}
% - - - - - - - - - - - - -
To gain insight on the DM mass distribution, we have to solve the appropriate equilibrium equations. We focus on dwarf galaxies, galaxies and clusters of galaxies assuming a negligible  visible matter contribution and a spherically symmetric halo of DM. These assumptions break down close to the galactic center, where the visible matter contribution is sizable. For a detailed description of the matter distribution it would be indeed  necessary to include the visible matter contribution. On the other hand, in dwarf galaxies, galaxies and clusters of galaxies the total contribution of the visible matter is negligible. Therefore, for the evaluation of the  the total mass and radius, as well as for estimating the matter distribution at large distances from the galactic centre, one can assume that the  dark  matter contribution is dominant.   
In this case, their nonrelativistic hydrostatic equilibrium  is approximately described by 
\begin{align}
\frac{d m}{dr} &= 4 \pi  \rho r^2\,, \label{eq:m} \\
\frac{d P}{dr} &=  -\frac{\rho m}{r^2} \label{eq:Newton_eq}\,,
\end{align}
where $P$ is the pressure and  $\rho$ is the DM matter density,  while $m(r)$ is the mass within the radius $r$. The  boundary conditions are
\be
\rho(0) = \rho_c \qquad m(0)=0\,,
\ee
where $\rho_c$ is the central DM density and we are assuming that there is no mass singularity at the  center. The solution of this Cauchy problem requires the knowledge of the EoS of DM, which relates pressure and matter density. For simplicity, we  consider a polytropic EoS 
\be\label{eq:polytrope}
P  = K \rho^{1+1/n}\,,
\ee
where $K$ is a constant and $n$ is called the poytropic index. Upon parametrizing the matter density as 
\be\label{eq:rhocpoly}
\rho= \rho_c \theta^n\,,
\ee
where  $\theta$ is a function of the radial coordinate,  the hydrostatic equilibrium equation can be written in a compact form as a Lane-Emden's differential equation
\be\label{eq:Lane-Emden}
\frac{1}{\xi^2}\frac{d}{d \xi}\left( \xi^2 \frac{d \theta}{d \xi}\right)= -\theta^n\,,
\ee
where we have rescaled the radius  as
$r = c_n \xi$, with 
\be 
c_n=\sqrt{\frac{(n+1)P_c}{4 \pi\rho_c^2}}=L_n\rho_c^{\frac{1-n}{2n}} \,,
\ee 
where 
\be
P_c = K \rho_c^{1+1/n}\,,
\ee
is the central pressure.

For polytropes there exists a simple  scaling relation between mass and radius, see for instance \cite{Shapiro-Teukolsky}, given by
\be\label{eq:scaling}
M = C_n R^\alpha\,, 
\ee
where \be\label{eq:alpha}
\alpha= \frac{3-n}{1-n} \,,
\ee
and
\be
C_n = -\frac{4 \pi}{L_n^{2n/(1-n)} \xi_1^{(1+n)/(1-n)}} \left.\frac{\partial\theta}{\partial \xi}\right\vert_{\xi_1}  \,,
\ee
with $\xi_1$ the rescaled stellar radius. Upon requiring that the mass of the structure does not decrease with increasing central densities, we have that $n \le 3$; moreover, since any hydrostatically stable configuration should have a matter density that decreases towards the surface, one has from Eq. \eqref{eq:rhocpoly} that  $n \geq 0$. Combining these two results,  for non relativistic structures described by one single polytropic EoS, it turns out that
\be
0\le n \le 3\,.
\ee
The value of $n$ determines whether the object is bound by cohesive forces (hereafter self-bound object) or it is gravitationally bound.  The two regimes are characterized by two different behaviors of the radius. In gravitationally bound objects the radius decreases with increasing mass, while in self-bound objects the radius increases with increasing mass. Then, we have that
\be
\begin{cases} 0\le n \le 1 & \text{self-bound objects}\\
1 \le n \le 3 & \text{gravitationally-bound objects}
\end{cases}
\ee
Since we find $M \sim R^{3.1}$, i.e. $\alpha > 3$, it follows that $ 0\le n \le 1 $, suggesting that DM is a form of self-interacting dark matter (SIDM), which seems to be  self-bound (see \cite{Tulin:2017ara} for a review). 
The SIDM model was introduced to solve (or alleviate) the core-cusp problem in the DM distribution as well as the missing satellite problems, see for instance~\cite{Spergel:1999mh}. 

\begin{figure}[t!]
    \centering
    \includegraphics[width=0.7\textwidth]{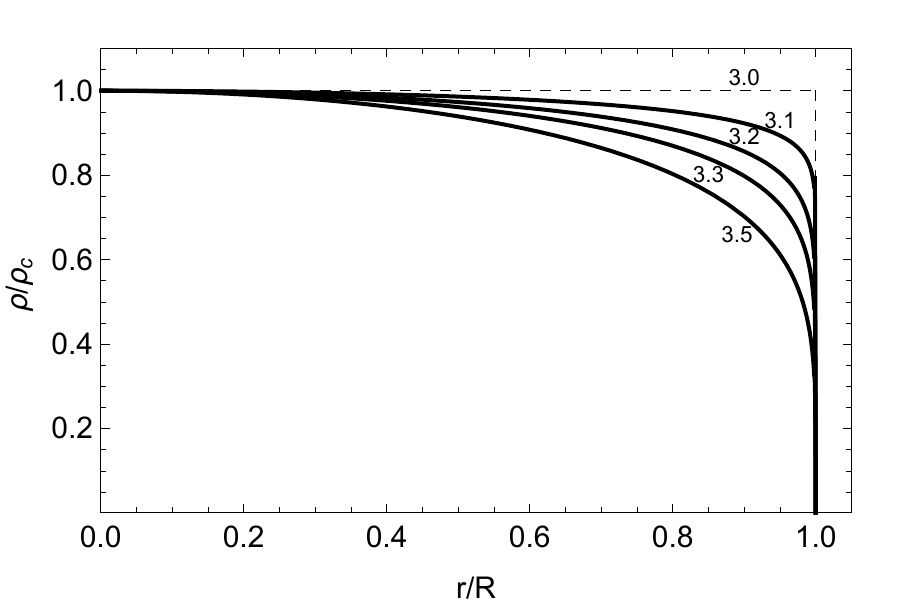}
    \caption{Dark matter density profiles 
    obtained solving the hydrostatic equilibrium  equations ~\eqref{eq:m} and \eqref{eq:Newton_eq} using a polytropic EoS, see Eq.~\eqref{eq:polytrope}. 
The radial distance has been     normalized distance to total radius of the galactic halo.     The different curves correspond to different  values of $\alpha$, see   Eq.~\eqref{eq:alpha}.}
    \label{fig:profiles}
\end{figure}

In Fig.~\ref{fig:profiles} we report a few mass density profiles obtained solving the hydrostatic equilibrium equations for different values of $\alpha$ close to the value obtained by the large scale structure fit in the mass-radius plot. The matter distribution is basically constant for $r<R $ and quickly decreases close to $R$; a behavior  typical of self-bound objects. Note that if DM is self-bound it means that diluted clumps of any size are possible.  If DM were the only form of matter,  this would imply that along the ``dark matter" line of Fig.~\ref{fig:2} one should find  different structures in between the large structures and the quantum mechanics boundary. In this case, the behavior would be similar to the one found along the ``atomic density" line. However, in the presence of visible matter it is not clear what happens to these small clumps of DM. If they exist, they will probably be captured by other visible structures.

The  inferred mass scale  ($< 0.1$ eV) matches the expected region of the neutrino masses, even though  a fermion of such mass scale can hardly be accepted as a DM candidate~\cite{Tremaine:1979we}. The coincidence is anyway surprising. Such a fermionic candidate would presumably represent a \emph{hot dark matter} component that cannot form halo structures around galaxies, being presumably relativistic. This picture could change if DM has a behavior similar to quark matter at high baryonic densities. In that case it is possible that fermionic matter forms strange stars~\cite{Alcock:1986hz}, a form of  self-bound and cold quark matter  expected to have  an  extremely small compressibility.   Indeed, these stars have a typical $M\propto R^3$ behavior. 

A different and somehow more conventional possibility is that  DM consists of  bosons with  a strong repulsive interaction.  Also in this case, thanks to the strong interaction, the compressibility would be extremely small. Bosons at low temperatures are expected to form a Bose-Einstein condensate (BEC); various realizations of DM as a BEC have been proposed~\cite{Ji:1994xh,Boehmer:2007um} even within a relativistic implementation, see  \cite{Bettoni:2013zma}.  In our case, in order to have a $M\propto R^3$ behavior one should have an interaction coupling which is strongly dependent on the matter density. 
Indeed if DM forms a superfluid gas with a point-like interaction it may be  described by the Gross-Pitaevskii equation, which gives  a pressure \be
p  = \frac{\rho^2 g}{2 m^2}\,,
\ee
where $g$ is the coupling constant. This means that if $g$ is constant, the GP equation gives a polytropic EoS with $n=1$. A density dependent interaction or some other modification of the Gross-Pitaevskii equation, see for instance \cite{Chavanis:2018pkx},  may produce a stiffer EoS. 

To briefly summarize  our results, we can say that  by assuming that DM is described at any scale by a single polytropic EoS it  emerges a picture of   the dark matter halo as a gas of  particles characterized by a small compressibility. This means that DM particles  are presumably self-bound. We will now apply this model to the rotation curves of galaxies.

% - - - - - - - - - - - - - - - - - -
\section{Rotation curves of galaxies}
\label{sec:rotation}
% - - - - - - - - - - - - - - - - - -

Astrophysical observations show a clear absence of a Keplerian fall in the rotation curves of disk galaxies: the radial profiles of the stars velocities $v(r)$ at large radii do not show the expected behavior proportional to $1/\sqrt{r}$ by visible matter observations. The observed velocity profile is almost constant at large radii, especially for the smaller and less luminous galaxies (see~\cite{bib:revsal} for a review). The simplest explanation is identified in the presence of an additional invisible dark mass component distributed in a spherically symmetric halo. The rotation velocity of stars is obtained assuming that the centrifugal force equates the gravitational pull, that is
\be\label{eq:centrifugal}
 v^2(r ) = \frac{G_N m_\text{tot}(r)}{r}\,,
\ee
where $m_\text{tot}(r)$ is the sum of the visible and dark masses  within the radius $r$.  Therefore, in order to exactly solve the problem, one should know the matter density profile  of dark and visible matter. In the previous section we approximately determined the DM distribution, neglecting visible matter. To approximately  take into account the contribution of visible matter, we assume that it contributes in quadrature to the total velocity, thus  the total radial velocity can be written as 
\be\label{eq:vstar} v^2(r) = v_{\text{d}}^2(r) + v_{\text{h}}^2(r) \,, \ee
where $v_\text{d}^2(r)$ is the visible disk contribution, whereas $v_\text{h}^2(r)$ is the DM halo contribution. For disk galaxies,  the visible  contribution has  an exponential  density profile \cite{bib:freeman}, thus the visible contribution to the stellar rotation velocity turns to be
\begin{equation}\label{eq:vdisk}
    v^2_\text{d}(x) = \frac12 \frac{G_N M_d}{R_d^3}r^2(I_0K_0-I_1K_1),
\end{equation}
where $M_d$ is the galaxy visible mass, related to the radius, $R_d$, through the density $\Sigma(r) = M_d/(2\pi R_D^2)e^{-r/R_d}$
and $I_n, K_n$ are the modified Bessel functions computed at $r/2R_d$. The  dark component velocity  contribution, assuming that DM is spherically distributed,  is computed as
\begin{equation}\label{eq:vhalo}
v_\text{h}^2(r) = \frac{G_N}{r}\int_0^r 4 \pi \rho_\text{h}(r') r'^2 dr'=  \frac{G_N M_\text{h}}{r} \frac{\int_0^{r/R_\text{h}} \rho_\text{h}(r') r'^2 dr'}{\int_0^1 \rho_\text{h}(r') r'^2 dr'} \,,   
\end{equation}
where $r'=r/R_h$ is the rescaled radial distance and $M_h$ and $R_h$ are the total mass and radius of the halo, respectively.  The halo mass distribution, $\rho_\text{h}(r)$ is  determined by solving Eqs.~\eqref{eq:m}  and \eqref{eq:Newton_eq} with the polytropic EoS in Eq.~\eqref{eq:polytrope} with $n\simeq 0.05$ corresponding to $\alpha=3.1$. For definiteness we will assume that $\alpha=3.1$, as from the fit of the mass-radius large scale structures.

\begin{figure}[tb!]
    \centering
    \includegraphics[width=0.7\columnwidth]{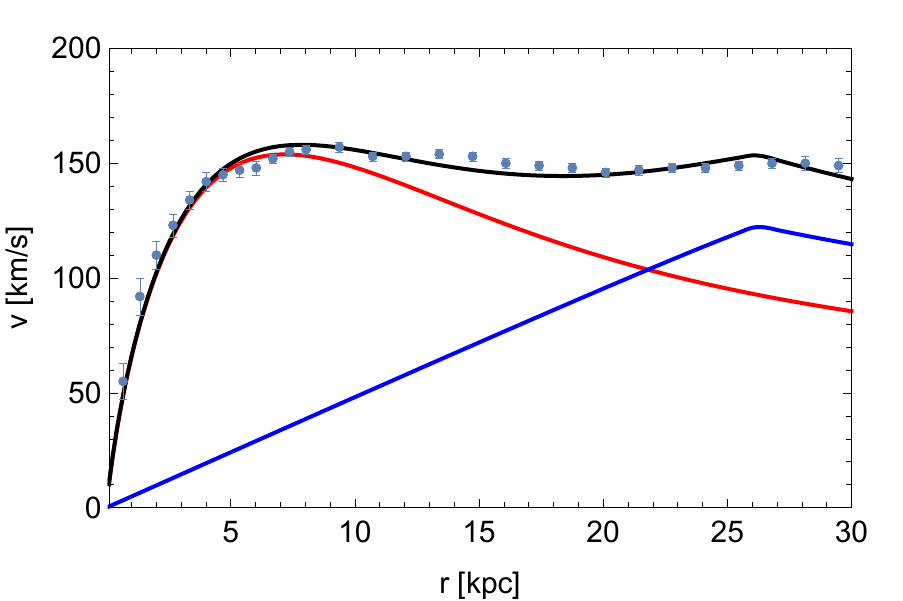}
    \caption{Rotation curve of the galaxy NGC3198. Data from \cite{1991A&A...244...27C,Daigle:2006fs,Gentile:2013tfa}, see also \cite{2015A&A...578A..13K}. The red curve represents the visible contribution, obtained with Eq.~\eqref{eq:vdisk}, while the blue curve represents the dark matter contribution, obtained with Eq.~\eqref{eq:vhalo}. The contribution of the dark matter halo is obtained using as   fit parameters  the  halo mass and radius; the  obtained values are $9.2 \times 10^{10}$ M$_{\odot}$ and $26.2$ kpc respectively.}
    \label{fig:ngc3198}
\end{figure}

Upon explicitly considering the dependence of the velocities on the macroscopic parameters, 
the total circular velocity can be written as
\be \label{eq:fit}
 v^2(r; M_\text{d}, R_\text{d}, M_\text{h}, R_\text{h}) = 
 v_\text{d}^2(r; M_d, R) +  v_\text{h}^2(r; M_\text{h}, R_\text{h}) 
 \ee
where the visible mass ans radii can  be estimated by observations, while we will use the halo mass and radius as fitting parameters.

The  halo mass and  radius depend on the particular  considered structure.  As a first application of our model, we consider in 
Fig.~\ref{fig:ngc3198} the rotation curve of the spiral galaxy NGC 3198~\cite{bib:ngc3198}.  Spiral galaxies typically show rotation curves which rapidly increase at short distances and then become flat far from the galactic center, meaning that they are expected to have  large and massive halos~\cite{Sofue:2000jx}. The rotation curves are considered as one of the best means to infer their mass~\cite{Sofue:2000jx}.
The spiral galaxy NGC 3198    exhibits  a steeply rising velocity profile close to the galactic centre and a wide flat velocity profile from $5$ to $30$ kpc.  The mass density distribution of this galaxy should be  approximately constant below $5$ kpc and then declining, reducing by about an order of magnitude at   $30$ kpc. The estimated visible mass of this galaxy is about $10^{10} M_\odot$, while the total mass at $30$ kpc is  $M \sim  5 \div 11 \times 10^{10} $ M$_\odot$, see for instance \cite{Daod_2019} and \cite{Kostov:2006vp}. In our approach the two fitting parameters are the halo mass and radius. We find that $M_\text{h} \simeq 9.2 \times 10^{10} M_\odot$, consistent with the estimate of \cite{Kostov:2006vp}. The fitted radius of the halo is instead $26.2$ kpc.  Given the fact that our model is just a simple polytropic EoS with two fitting parameters, the agreement with the estimated halo mass and the with the observed velocity profile is quite promising.

\begin{figure*}[t!]
    \centering
    \includegraphics[width=1.\textwidth]{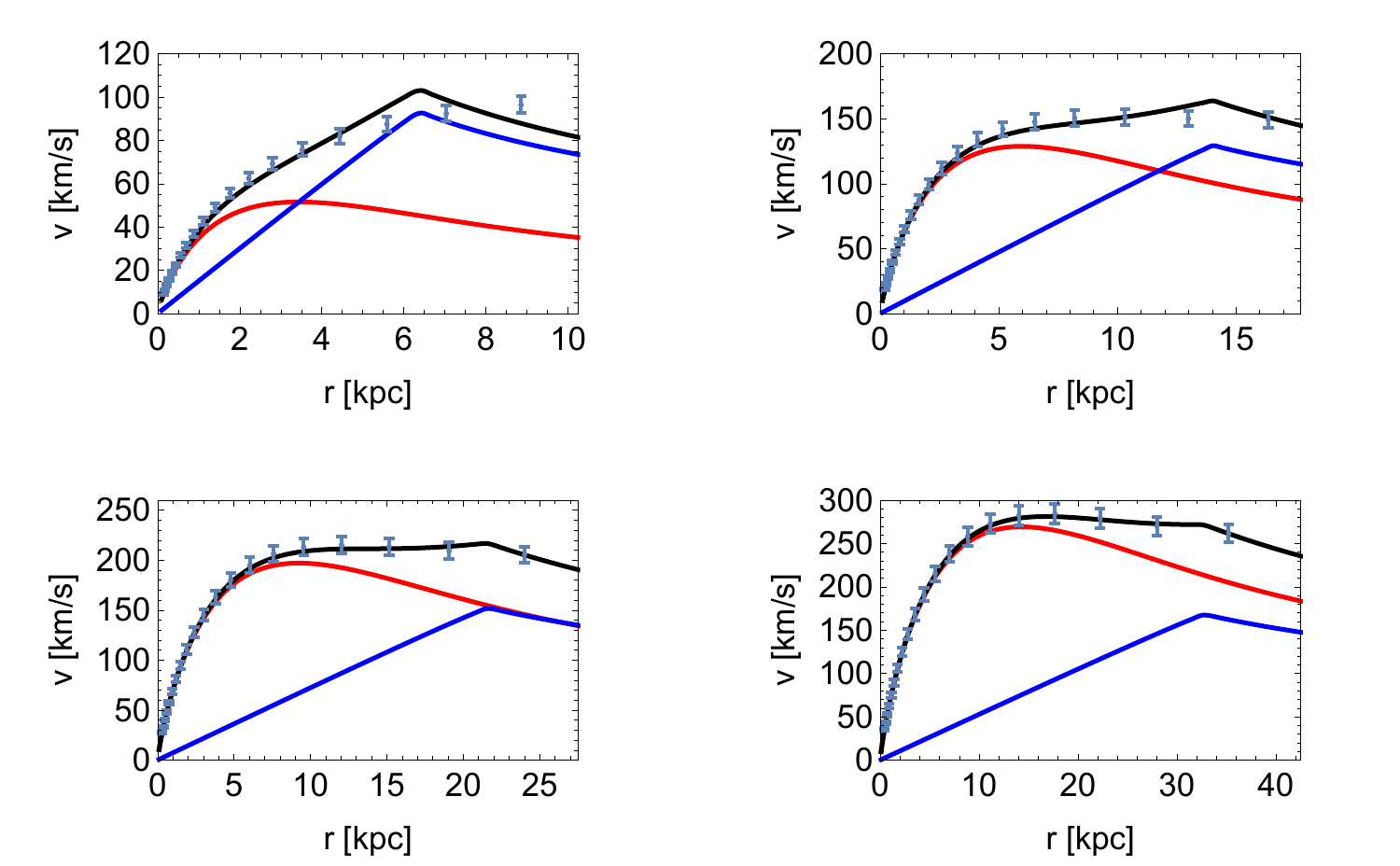}
    \caption{Rotation curves for various galaxies. The  dots are obtained by  modelling  the corresponding sample uncertainty by so-called Universal Rotation Curves (URC). 
    The red curve represents the visible contribution, obtained with Eq.~\eqref{eq:vdisk}, while the blue curve represents the dark matter contribution, obtained with Eq.~\eqref{eq:vhalo}.
     The dark matter contributions are obtained  by fitting the  masses and the radii of the dark matter  halos. 
     From left to right, top to bottom, the estimated visible masses are $M_d\simeq (10^{9.4}, 10^{10.4}, 10^{11.0}, 10^{11.5} ) M_\odot$, respectively, see~\cite{bib:urc1,bib:urc2}. 
     With our model, the fitting values of the  halo masses and radii are $(1.61, 5.46, 11.7, 21.6) \times 10^{10}$ M$_\odot$ and $(7.16, 14.1, 31.6, 32.8)$ kpc, respectively.}
    \label{fig:samples}
\end{figure*} 

For this reason we extended the analysis to  rotation curves of different galaxies~\cite{bib:urc1,bib:urc2} whose estimated visible masses   range from $M_d \sim 10^{8.5} M_\odot$ to $M_d \sim 10^{11.5} M_\odot$.   
In Fig. \ref{fig:samples} we report some results obtained  fitting the values of $M_\text{h}$ and $R_\text{h}$   for galaxies of different sizes: the red and the blue curves represent the visible and the halo components, respectively.  
In spite of the simplification of the model, it can appropriately reproduce the rotation velocity profiles with the appropriate choice of a suitable halo size and mass.
The sharp discontinuity in the halo component, visible especially for lighter galaxies, is due to the fast transition of the DM distribution from a constant density to zero. In other words, the kink in the rotation curves is determined by the fact that in our model DM is self bound.   In a more refined description, one should  determine $\rho_\text{h}(r)$ including in the hydrostatic equilibrium equation~\eqref{eq:Newton_eq} the contribution of the visible matter, which would certainly reduce the effect of the kink in the DM distribution. 

% - - - - - - - - - - - - - - - - - - - -
\section{Conclusions}
\label{sec:conclusions}
% - - - - - - - - - - - - - - - - - - - -

The standard cold DM model made of WIMPs exhibits several issues not yet completely solved, as the \emph{cusp-core}, the \emph{diversity}, the \emph{missing galaxy satellites}, and the \emph{too-big-to-fail} problems \cite{bib:tulin, bib:bullock}, all related to the galactic structures  with  DM cores. Many different methods have been introduced to address these problems: among them self-interacting dark matter models seem to be  viable candidates to  solve or at least mitigate them~\cite{Spergel:1999mh}. In these approaches the DM mass and its self-interactions are introduced by hand and  are properly tuned to produce the  desired DM distribution in galaxies. 

We have presented a novel semi-quantitative approach to DM based on the analogy between the observed behavior of matter at different scales, revisiting and extending the mass-length diagram proposed in \cite{bib:Carr}. In terms of fundamental parameters, such as $m_p$, $m_e$, $\alpha_e$, $\alpha_G$,  $\alpha_W$ and $\alpha_s$ we have given an approximate justification of the mass-radius relationship for  most of the observed structures in the Universe. In particular, we have shown that from existing measurements of masses and radii of  stars and planets one may infer the mass of the electron by properly extrapolations at small scales. As a matter of fact, the intersection point of the ``atomic density" line with the quantum mechanics boundary, see Fig~\ref{fig:2},  corresponds to a mass scale of about $0.4$ MeV, close to the electron mass. Similarly, from masses and radii observations of neutron stars  and nuclear matter, one may infer a mass scale of about $600$ MeV, close to  the proton mass.  Using the same argument, but applied to galaxies, we have inferred that if the DM halos consists of a single type of  elementary particle, its mass  should be of the order of $m_\chi \simeq 1$--$100$ meV. Moreover, from the slope of the line connecting the various considered galaxies, it seems that the dark-matter EoS can be approximated by a polytrope with a small polytropic index, characteristic of strongly self-interacting matter.
 
We do not put forward any hypothesis about the origin of this new mass scale and the connection with extensions of the  Standard Model of particle physics. We only observe that if the mass-radius of very large structures is determined by a dark-matter particle its mass could be of the order of $1$--$100$ meV. Moreover, given the low compressibility of the EoS that describes DM in these galaxies, the interaction strength should be 
sizeable, possibly comparable with that of nuclear matter, see  \cite{Tulin:2017ara} for a similar result.  

Using the obtained EoS we have shown that the mass distribution in galaxies is such that  their observed rotation velocities  seem to be qualitatively well described. Given the simplicity of the model, in which the visible matter is approximately included and DM at any scale is described by one single polytrope, the obtained results are quite promising. 
Since the proposed  DM is self-bound, it has a rather uniform mass distribution, which more or less abruptly ends (depending on the polytropic index used). 

The present analysis can be improved considering different levels of complexity. The visible and dark matters should be described together with a common Tolman-Oppenheimer-Volkoff equation~\cite{Tolman, Oppenheimer-Volkoff}, this would yield an  accurate total mass distribution. In analogy with the atomic and nuclear lines, another possibility is to study the smearing of the galactic distribution around the ``dark matter" line, due to internal dynamics and formation mechanisms.

Finally, it is worth pointing out that the star distribution makes an exception in the mass-radius diagram:  the presence of several interaction processes produces a sizable deviation with respect to the the atomic alignment, see Fig.~\ref{fig:2}. A similar behaviour is observable for the globular clusters that deviate on the left of the ``dark matter" line. This seems to suggest that there could be a transition region connecting stars to galaxies passing through globular clusters. In other words, in this region the Standard Model interaction and the DM interaction may have a comparable strength. Such behavior could be similar to the  one observed on the left of the  stars' distribution. Although neutron stars, white dwarfs and standard stars form three distinct clusters,   we observe  a kind of continuous mass distribution linking these three clusters. The white dwarfs  are indeed located just in between the neutron stars and the red dwarfs, corresponding to the region in which  nuclear, electromagnetic and exchange interactions are at play.

In summary, the proposed semi-quantitative method of clustering structures in the mass-radius diagram seems to be an excellent tool for global descriptions linking the micro- and macro-physics shedding light on how fundamental interactions work, including the DM contribution.

\section*{Acknowledgements}

We would like to thank Paolo Salucci for providing the data points of URCs. We would also like to thank  Carlos Pen\~a Garay, Daniele Montanino and Nicolao Fornengo for 
useful and relevant discussions.

\bibliography{bibfile}

\end{document}